\documentclass[sigconf]{acmart}
\acmConference[EASE 2026]{The 30th International Conference on Evaluation and Assessment in Software Engineering}{9–12 June, 2026}{Glasgow, Scotland, United Kingdom}
\acmDOI{}
\acmISBN{}

\AtBeginDocument{%
  }
\usepackage{enumitem}
\usepackage{multirow}
\usepackage{listings}
\usepackage{xcolor}
\usepackage{tabularx}
\usepackage{array}
\usepackage{colortbl}
\newcommand{\best}[1]{\cellcolor{gray!15}\textbf{#1}}

\usepackage{mdframed}
\usepackage{caption}
\mdfdefinestyle{graybox}{
    backgroundcolor=gray!12,
    linecolor=black!60,
    linewidth=0.0pt,
    roundcorner=3pt,
    skipabove=4pt,
    skipbelow=2pt,
    innertopmargin=4pt,
    innerbottommargin=4pt,
    innerleftmargin=4pt,
    innerrightmargin=4pt
}

\definecolor{lstbg}{HTML}{FBFBFD}
\definecolor{lstframe}{HTML}{B8C0CC}
\definecolor{lstkw}{HTML}{1D4ED8}      
\definecolor{lstcom}{HTML}{16A34A}     
\definecolor{lststr}{HTML}{B45309}     
\definecolor{lstnum}{HTML}{6B7280}     

\begin{document}

\title{Leveraging LLMs for Multi-File DSL Code Generation: An Industrial Case Study}


\author{Sivajeet Chand}
\authornote{Both authors contributed equally to this research.}
\affiliation{%
  \institution{Technical University of Munich}
  \city{Munich}
  \country{Germany}
}
\email{sivajeet.chand@tum.de}

\author{Kevin Nguyen}
\authornotemark[1]

\affiliation{%
  \institution{BMW Group}
  \city{Munich}
  \country{Germany}
}
\email{kevin.k.nguyen@bmw.de}

\author{Peter Kuntz}
\affiliation{%
  \institution{BMW Group}
  \city{Munich}
  \country{Germany}
}
\email{peter.kuntz@bmw.de}

\author{Alexander Pretschner}
\affiliation{%
 \institution{Technical University of Munich}
 \city{Munich}
 \country{Germany}
  }
\email{alexander.pretschner@tum.de}

\renewcommand{\shortauthors}{Sivajeet et al.}

\begin{abstract}
  Large language models (LLMs) perform strongly on general-purpose code generation, yet their applicability to enterprise domain-specific languages (DSLs) remains underexplored, especially for repository-scale change generation spanning multiple files and folder structures from a single natural-language (NL) instruction. We report an industrial case study at BMW that adapts code-oriented LLMs to generate and modify project-root DSL artifacts for an Xtext-based DSL that drives downstream Java/TypeScript code generation. We develop an end-to-end pipeline for dataset construction, multi-file task representation, model adaptation, and evaluation. We encode DSL folder hierarchies as structured, path-preserving JSON, allowing single-response generation at repository scale and learning cross-file dependencies. We evaluate two instruction-tuned code LLMs (Qwen2.5-Coder and DeepSeek-Coder, 7B) under three configurations: baseline prompting, one-shot in-context learning, and parameter-efficient fine-tuning (QLoRA). Beyond standard similarity metrics, we introduce task-specific measures that assess edit correctness and repository structural fidelity. Fine-tuning yields the most significant gains across models and metrics, achieving high exact-match accuracy, substantial edit similarity, and structural fidelity of 1.00 on our held-out set for multi-file outputs. At the same time, one-shot in-context learning provides smaller but consistent improvements over baseline prompting. We further validate practical utility via an expert developer survey and an execution-based check using the existing code generator.
\end{abstract}




\maketitle

\section{Introduction}
\label{sec:introduction}

Transformer-based LLMs have become effective assistants for code-related tasks in general-purpose languages (GPLs) \cite{11029737, 10.1145/3597503.3639219, pandey2025design}, but there is still limited empirical evidence on their effectiveness for enterprise DSLs \cite{gu2025}, particularly for repository-scale generation involving multiple files and folder structures from a single NL instruction. In industrial settings, DSLs are frequently used as the authoritative source from which generators produce downstream artifacts (e.g., Java/TypeScript), meaning that small errors in the DSL can break compilation or generation and directly impact engineering workflows.

This paper reports an industrial case study with BMW Financial Services, where a custom Xtext-based DSL \cite{efftinge2006} is used to generate boilerplate and configuration artifacts across multiple regional markets. The DSL serves as an abstraction layer that accelerates development, but authoring and evolving the DSL remains manual and requires substantial expertise—particularly when creating or adapting market configurations that involve coordinated changes across multiple folders and files. Our goal is to let developers (and potentially non-technical stakeholders) specify high-level changes in NL and automatically obtain a consistent, project-root set of DSL artifacts consumable by the existing code generator.

\begin{figure*}
    \centering
    \includegraphics[width=0.71\textwidth]{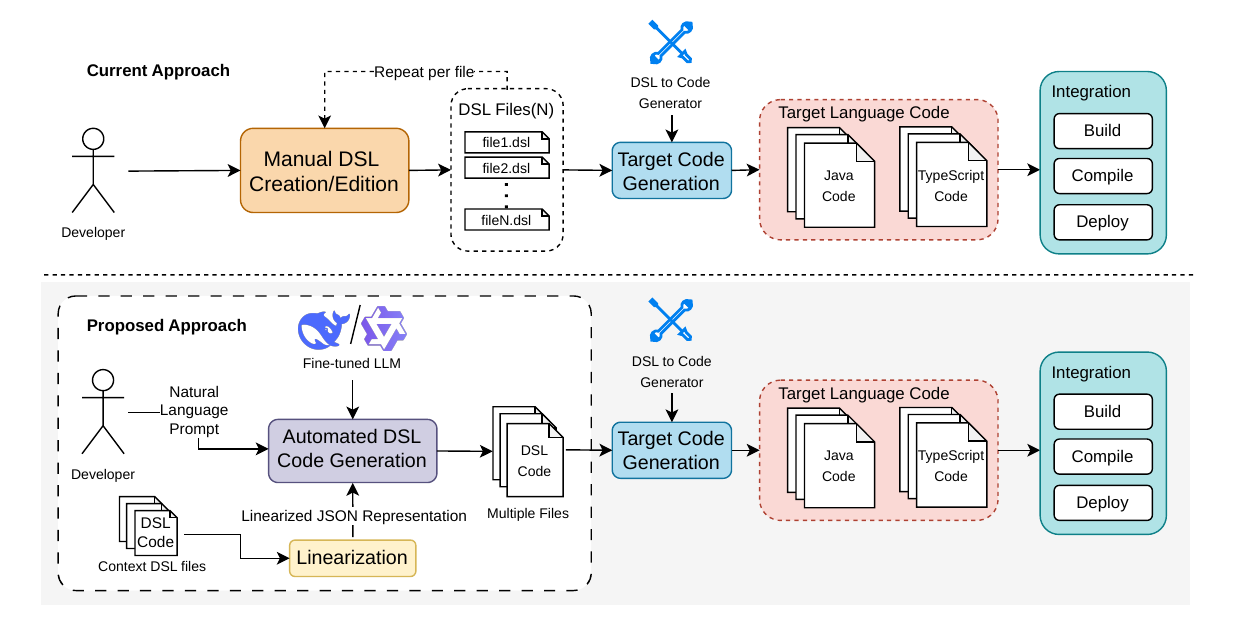}
    \caption{Overview of current workflow vs the proposed workflow}
    \label{fig:DPR}
\end{figure*}

A key challenge is repository-scale generation: one instruction can require coordinated edits across multiple DSL files with cross-file dependencies (e.g., adding an attribute in one layer implies updates in related layers). To fit within context-window limits and enable objective evaluation, we linearize the entire DSL folder hierarchy into a single JSON document, using file paths as keys and file contents as values. This lets the model operate at project-root level in one generation step and supports systematic comparison of original context, target state, and model output.

Figure~\ref{fig:DPR} illustrates the industrial workflow and how the proposed approach changes it: in the current (“as-is”) process, developers manually create/modify DSL files file-by-file; in the proposed (“to-be”) process, a NL instruction is translated into a complete, consistent set of DSL files, which is then passed to the existing DSL generator to produce downstream artifacts. The critical success criterion is therefore not only textual similarity, but that the generated DSL artifacts preserve repository structure and satisfy the DSL toolchain requirements.

To investigate feasibility and cost–benefit trade-offs of adapting LLMs, we evaluate two instruction-tuned LLMs (Qwen2.5-Coder and DeepSeek-Coder, 7B) under three configurations: (i) baseline zero-shot prompting, (ii) one-shot in-context learning (ICL) with DSL rules and an example \cite{brown2020}, and (iii) parameter-efficient fine-tuning using QLoRA \cite{hu2021}. Because generic similarity metrics alone are insufficient for repository-scale DSL changes, we combine standard measures (e.g., exact match, BLEU, JSON validity) with task-specific metrics that focus on edit correctness and structural fidelity of the generated folder/file tree. We further validate practical utility through execution-oriented checks using an existing DSL generator and an expert review by senior developers familiar with the DSL.

\textbf{This paper makes the following contributions}:
\begin{itemize}[topsep=0pt, partopsep=0pt, itemsep=0pt, parsep=0pt]
    \item An industrial BMW case study on repository-scale NL-driven generation and modification of Xtext-based DSL artifacts.
    \item A structured linearization that encodes multi-file DSL repositories into a single input/output for project-root generation and consistent cross-file updates.
    \item A task-specific evaluation framework that complements standard metrics with change-focused similarity and structural fidelity measures for multi-file DSL changes. 
    \item An empirical comparison of zero-shot, one-shot ICL, and parameter-efficient fine-tuning (QLoRA) on this industrial DSL task, including execution- and expert-based validation.
\end{itemize}

We evaluate these contributions via three research questions:

  \textbf{RQ1: \textit{How accurately can LLMs generate repository-scale DSL change-sets (files, folders, and contents) from a natural-language instruction?}} \\This research question establishes the feasibility of applying LLMs to the proposed task.

 \textbf{RQ2: \textit{To what extent do prompting and fine-tuning improve LLM performance on this task?}} \\This question assesses the incremental gains of adaptation techniques over the baseline, weighing prompt engineering and fine-tuning effort against improvements across validity, structural correctness, and semantic similarity.

  \textbf{RQ3: \textit{How useful are the generated DSL change-sets to developers who maintain these DSL repositories in practice?}} \\This question examines whether LLM outputs provide workflow value by reducing authoring time and correction effort while maintaining acceptable quality.

\section{Background}\label{Background}
DSLs raise the level of abstraction by providing constructs tailored to a particular domain, enabling more concise specifications than GPLs and supporting automation via generators \cite{voelter2013,van2000,mernik2005,fowler2010}. In model-driven settings, DSL artifacts are commonly transformed into GPL code, which reduces boilerplate editing effort and can improve consistency and maintainability.

\begin{lstlisting}[language=Java, caption={Anonymized DSL snippet for server-side configuration}, label={lst:dsl_server}]
entity abstract FinanceProductBase extends FinanceProduct {
    attribute1: AttributeType16
    attribute2: AttributeType17
    attribute3: AttributeType8
    attribute4: AttributeType9
    attribute5: AttributeType9
    attribute6: AttributeType6
}
entity ProductTypeA extends FinanceProductBase {
    attribute7: AttributeType18
    attribute8 = FinanceProductTypeModule::type_a
    attribute9 = CategoryType::loan
}
\end{lstlisting}

As shown in Figure~\ref{fig:DPR}, in our industrial setting, the DSL is implemented with Eclipse Xtext \cite{efftinge2006,zhang2025xtext}. Xtext grammars define the DSL syntax and allow generation of parsers and abstract syntax representations, while a custom generator translates DSL files into Java and TypeScript artifacts for different subsystems. This workflow is effective, but it still relies on developers to manually author and update many DSL files, often across nested project structures. Our work targets automating this root-folder / multi-file DSL authoring step from NL instructions while maintaining generator compatibility.

\begin{lstlisting}[language=Java, caption={Anonymized Java Code Snippet for Server Configuration}, label={lst:java_server}]
@RegionSpecific(Region.MARKET)
public abstract class FinanceProductBase extends FinanceProduct 
{
    public static final String PROPERTY_ATTRIBUTE_1 = "attribute1";
    public static final String PROPERTY_ATTRIBUTE_2 = "attribute2";
    public static final String PROPERTY_ATTRIBUTE_3 = "attribute3";
    public static final String PROPERTY_ATTRIBUTE_4 = "attribute4";
    public static final String PROPERTY_ATTRIBUTE_5 = "attribute5";
    public static final String PROPERTY_ATTRIBUTE_6 = "attribute6";

    private AttributeType16 attribute1;
    private AttributeType17 attribute2;
    private AttributeType9 attribute3;
    private AttributeType9 attribute4;
    private AttributeType9 attribute5;
    private AttributeType6 attribute6;

    public FinanceProductTypeA(AttributeType18 attribute7,
                               FinanceProductTypeModule attribute8) {
        super(attribute7, attribute8);
        setAttributeType25(new AttributeType25());
    }
    public AttributeType16 getAttribute1() {
        return attribute1;
    }
    public void setAttribute1(AttributeType16 attribute1) {
        this.attribute1 = attribute1;
    }
    public FinanceProductTypeA attribute1(AttributeType16 attribute1) {
        this.attribute1 = attribute1;
        return this;
    }
}
\end{lstlisting}

Listing~\ref{lst:dsl_server} illustrates an anonymized and simplified excerpt of our Xtext-based DSL. The core construct is an \texttt{entity}, which supports inheritance and declares domain attributes at a high level of abstraction. In contrast, the corresponding generated GPL artifacts are considerably more verbose \cite{voelter2011}: Listing~\ref{lst:java_server} shows a representative Java excerpt produced from the DSL, including boilerplate such as constants, field declarations, constructors, and accessor methods. This highlights a key advantage of using DSLs: editing the DSL files is more efficient and less error-prone than directly modifying the Java codebase. Moreover, this conciseness is important for NL-to-DSL generation: compared to generating Java/TypeScript directly, the DSL captures the same domain intent with fewer tokens, making it feasible for an LLM to generate complete market configurations (including multiple interdependent files and folders) within context-window limits, while reducing the syntactic search space due to the DSL’s more regular grammar.

LLM-based code generation is predominantly built on the Transformer architecture \cite{vaswani2017}. Code-oriented LLMs benefit from pretraining on large corpora and can be adapted to specialized tasks via fine-tuning \cite{chen2021,brown2020}. However, enterprise DSLs typically have limited training data and strict grammars, making reliability and structural validity central concerns. To reduce fine-tuning cost, parameter-efficient fine-tuning (PEFT) methods such as LoRA and QLoRA are widely used \cite{hu2021,dettmers2023,xu2023}. As an alternative or complement, prompting / in-context learning (ICL) can elicit task behavior without weight updates \cite{brown2020,wei2022,reynolds2021,zhao2021,li2023,dong2024,sahoo2025}, but tends to be less stable for tasks requiring strict syntax and domain conventions, particularly under long-context and multi-file constraints \cite{mosbach2023,liu2022}.

Evaluating NL-to-code systems remains non-trivial. Text-overlap metrics such as BLEU are known to correlate weakly with semantic correctness \cite{papineni2002}. Code-specific variants (e.g., CodeBLEU) incorporate syntax/semantic signals, but are primarily designed for GPLs \cite{ren2020}. For DSL generation, functional unit-test execution (e.g., HumanEval-style evaluation) is often not directly applicable because the DSL artifacts are typically inputs to generators rather than executable programs \cite{chen2025}. Accordingly, DSL generation evaluation commonly combines: (i) syntactic validity (parse/grammar conformance), (ii) toolchain acceptance (generator/build success), and (iii) expert review of structural fidelity and maintainability \cite{paul2024}.

\section{Related Work}\label{relatedwork}
Early work on translating NL into formal representations is rooted in semantic parsing and program synthesis \cite{berant2013,dong2016}. A key line of research introduced grammar-constrained decoding and AST-based generation to increase syntactic validity in structured outputs \cite{yin2017,rabinovich2017}. These ideas are directly relevant to NL-to-DSL generation because DSLs are typically specified by explicit grammars and downstream generators are sensitive to syntactic errors.

Recent work focuses on specifically making LLM outputs for DSL-like formats structurally correct and parseable. Constraint-based and grammar-aware generation methods aim to guide decoding so that outputs remain machine-consumable \cite{wang2024}. Practical systems such as DSL-Xpert combine grammar prompting and example-driven prompting to improve reliability of NL-to-DSL translation \cite{lamas2024}. Other research explores using DSLs as intermediates for niche code generation \cite{kogler2025}. Empirical studies also report that general LLMs often underperform on DSLs without adaptation, motivating fine-tuning and stronger constraints \cite{gu2025,joel2024,bassamzadeh2024}.

Despite this progress, existing research provides limited evidence on project-level, multi-file DSL generation where the required output spans nested folders and multiple mutually dependent DSL artifacts, and where success is determined by compatibility with an existing industrial generator and build pipeline. Our paper addresses this gap by studying (i) fine-tuning of modern code LLMs for a proprietary Xtext-based DSL and (ii) comparison to prompting baselines, under an evaluation centered on syntactic validity and toolchain acceptance for root-folder/multi-file outputs.

\section{Methodology}\label{sec:methodology}

\begin{figure}[t]
  \centering
  \includegraphics[width=0.80\columnwidth]{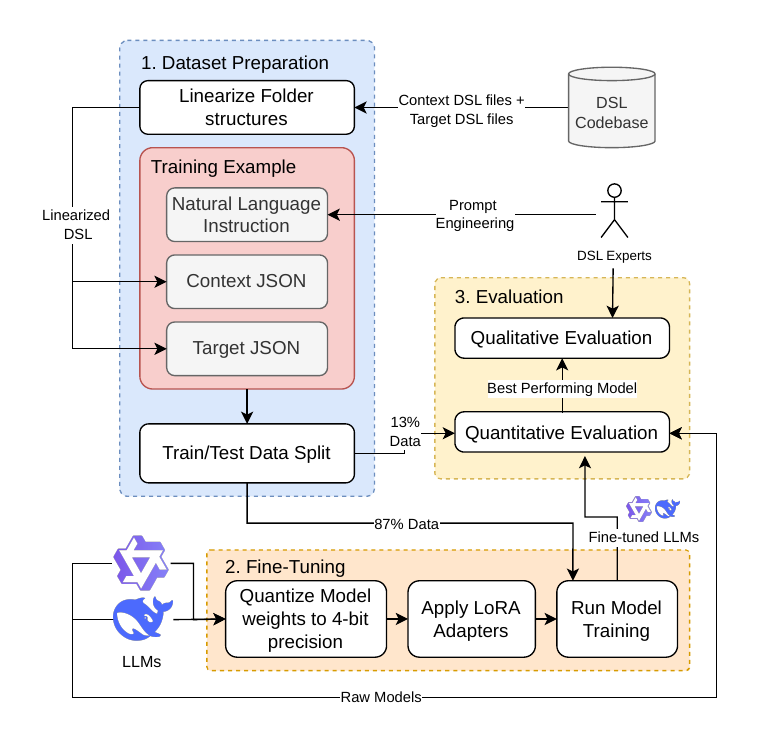}
  \caption{End-to-end workflow}
  \label{fig:data_workflow}
\end{figure}

We study NL-to-DSL generation at the root-folder level: given (i) a NL instruction and (ii) the current state of a market-specific DSL project (multiple folders/files), the system outputs the complete updated project state applying the requested change. The use case is that developers or non-technical stakeholders describe market creation or modifications in plain English, and the system generates or edits the corresponding DSL codebase across dependent files. Figure~\ref{fig:data_workflow} summarizes the workflow from data preparation to training and evaluation.

\subsection{Data preparation and task representation}\label{sec:data}
The DSL is implemented in Eclipse Xtext and defines market-specific project structures across multiple subsystems (e.g., server/UI). The key challenge is capturing \textbf{cross-file} and \textbf{cross-folder} dependencies so that a single instruction yields consistent multi-file updates.

\textit{Linearized project context and task.} For each market, we serialize the DSL folder hierarchy into a single structured JSON document: folders are nested JSON objects, files are nested keys, and file contents are stored as strings with newline characters preserved. This enables \textbf{root-folder-level} editing: given an instruction and a project snapshot (context JSON), the model generates the fully updated snapshot (target JSON), including all required cross-file changes (e.g., dependent edits and import updates).

\textit{Example structure and operations.} Each example is an (instruction, context, target) triple, where the instruction is manually authored with DSL experts and context/target are the before/after project snapshots. The dataset covers three operation families: (i) \textbf{create} a new market (often derived from an existing one), (ii) \textbf{add} features (attributes or finance products), and (iii) \textbf{delete} features (attributes or finance products). These operations are dependency-heavy: changes in one file can require consistent updates in other files, including adding/removing imports.

\textit{Minimal-context variants.} In addition to full-context examples, we include minimal-context variants of the same underlying tasks: the instruction is unchanged, but the context JSON retains only the most relevant files/folders (and simplified context files where applicable) to reduce noise from unrelated artifacts.

\textit{Dataset size, composition, and constraints.} We manually constructed the dataset (no public dataset exists for this proprietary DSL). It contains 774 training examples and 105 held-out evaluation examples. Of the training set, 309/774 (40\%) examples are minimal-context variants and 465/774 (60\%) are full-context. Across all training examples, average context/target sizes are 196/199 LOC with $\sim$6.1 files and $\sim$5.6 folders per example; minimal-context examples average $\approx$133/134 LOC with $\approx$4.3 files and $\approx$4.3 folders, whereas full-context examples average $\approx$239/243 LOC with $\approx$7.4 files and $\approx$6.5 folders.

\textit{Data split and leakage control.} Because minimal-context variants share the same instruction as their corresponding full-context tasks, we prevent leakage by assigning all variants derived from the same underlying market change to the same split.

\textit{SFT formatting and masking.} We convert each triple into an instruction-following format for instruction-tuned causal LLMs, tokenize using the base-model tokenizer, and mask labels so loss is computed only on the \emph{target output} (prefix masked with -100). We run prefix-alignment checks to prevent truncation or chat-template misalignment, store metadata (token counts and rendered prompts/targets), and export the processed data as a HuggingFace dataset.

\subsection{Model selection}\label{sec:models}
We fine-tune two open-source, instruction-tuned code LLMs: \\ \texttt{Qwen-2.5-Coder-7B-Instruct} \cite{hui2024} and
\texttt{DeepSeek-6.7B-Coder -Instruct} \cite{guo2024}.
We selected 7 billion parameter models as a balance between (a) capacity to model strict DSL syntax and multi-file dependencies and (b) resource/cost constraints. Smaller variants (1.5B/3B) were considered but risk underfitting for multi-file edits; larger variants (14B/32B) were excluded due to budget and increased overfitting risk given $<1000$ training samples. Both models have adequate context to include instruction, a full linearized market context, and generate the fully updated JSON without truncation.

\subsection{Prompting Baselines}
To benchmark fine-tuning against in-context learning, we evaluate two prompting configurations on the unfine-tuned base models. In \emph{zero-shot} setting, the prompt includes only the raw user instruction and the linearized context JSON, requiring the model to infer the edits without further guidance. In \emph{one-shot} setting, we add a compact summary of DSL grammar rules and a single instruction–output demonstration showing the expected JSON format and typical cross-file edits, to steer the model toward syntactically valid DSL and correct project-level structure. By contrast, prompts for supervised fine-tuning include only the raw instruction (plus context JSON), since syntactic regularities and formatting conventions are learned from training data rather than provided at inference time.

\subsection{Fine-tuning with QLoRA}\label{sec:finetune}
Given the limited dataset size, we use parameter-efficient fine-tuning rather than full fine-tuning to reduce overfitting. Specifically, we apply QLoRA \cite{dettmers2023}: base weights are quantized to 4-bit using \texttt{bitsandbytes} \cite{dettmers2022}, while computation uses \texttt{float16} \cite{micikevicius2018}. We additionally enable FlashAttention \cite{dao2022,dao2023} where supported. This significantly reduces the memory required to store the model weights.

\textit{LoRA configuration.}
We use rank $r=16$, scaling $\texttt{lora\_alpha}=32$, dropout $0.05$, $\texttt{bias}=\texttt{none}$, and $\texttt{task\_type}=\texttt{CAUSAL\_LM}$ \cite{hu2021,dettmers2023,yan2025}. Adapters are applied to attention \texttt{q\_proj} and \texttt{v\_proj}; \texttt{k\_proj} and \texttt{o\_proj} remain frozen to keep the trainable parameter count low while preserving adaptation capacity.

\textit{Training.} Training runs are executed on AWS SageMaker using a custom Python script built on HuggingFace Transformers \cite{wolf2020} and PyTorch. We monitor training via SageMaker Studio logs and select checkpoints by evaluation loss. To reduce memory usage, we enable gradient checkpointing \cite{chen2016} and use gradient accumulation \cite{lamypoirier2021}. All final training runs use \texttt{ml.g5.xlarge} instances (minimum viable GPU memory for end-to-end training).

\textit{Hyperparameters.}
For both models, we use: per-device batch size $=1$, gradient accumulation steps $=8$ (effective batch size $=8$), learning rate $1\mathrm{e}{-4}$, optimizer \texttt{adamw\_bnb\_8bit}, 2 epochs, warmup steps $=10$, evaluation every 10 steps, and checkpoint saving every 10 steps. The best checkpoint is persisted to S3.

\subsection{Evaluation}\label{sec:evaluation}
We evaluate (i) the two fine-tuned models and (ii) baselines on the same held-out set. For each base model, we compare:
\textbf{zero-shot} prompting, \textbf{one-shot} prompting (with grammar summary + one demonstration), and the \textbf{fine-tuned} variant. The best-performing fine-tuned model is then used for deeper qualitative review and a prototype demonstration.

\textit{Quantitative metrics.}
We report:
\begin{itemize}[style=nextline,leftmargin=8pt,labelwidth=0pt, topsep=0pt ]
    \item \textbf{Exact match} (binary) compares the generated DSL output with the expected output and yields a 0 or 1 score depending on whether they are identical or not \cite{rabinovich2017,soliman2022}.
    \item \textbf{BLEU} scores a candidate text 'c' against one or more reference texts 'r' by combining modified n-gram precision with a brevity penalty \cite{papineni2002}.
    \item \textbf{Valid JSON} check to ensure outputs are parseable as the expected linearized structure.
    \item \textbf{Custom DSL metrics} targeting (a) change similarity and (b) structural fidelity .
\end{itemize}

\paragraph{Change similarity.}
Standard metrics are dominated by unchanged context when edits are small, so we score \emph{only the expected changes}. For each example, we flatten the context/target/predicted JSON into leaf key–value pairs, use Python \texttt{SequenceMatcher} to compute a line-level difference between context and target, and compare the corresponding changed lines in target vs. prediction at token level. We then compute the normalized token-level Levenshtein error per changed line:
\begin{align}
e_i = \frac{d_\text{lev}(t_i, p_i)}{\max(|t_i|, |p_i|, 1)} \label{eq:levenshtein}
\end{align}

The errors E are accumulated, resulting in a raw similarity score between 0 and 1 for each changed line.

This similarity score is then exponentiated with a parameter $\alpha = 5.0$ to magnify the error penalties for larger inaccuracies, resulting in the final similarity score per line:
\begin{equation}\label{eq:similarities}
E=\sum_i e_i\quad,\quad
S_{\text{raw}}=1-E\quad,\quad
S_{\text{line}}=S_{\text{raw}}^{\alpha}
\end{equation}

To compute the overall average change similarity score of the model, a log-weighted aggregation is applied. First, for each changed line \(c \in \mathcal{C}\), a weight is computed based on its size. This ensures that larger changes contribute more to the final similarity score, while in the same time the logarithm prevents very large changes from dominating the whole score:
\begin{align}
w(c) = \min\Big(\log\bigl(1 + \text{len}(v_{\text{true},c})\bigr), w_\text{max}\Big)
\label{eq:line_weight}
\end{align}

Here \(\text{len}(v_{\text{true},c})\) denotes the size of the true line content for change \(c\), and \(w_\text{max} = 20.0\) is the maximum cap for the weight. 

Finally, the line-level token similarity score \(S_\text{line}(c)\), computed using a combination of line-level diffing,
token-level alignment, and Levenshtein distance is combined with the line weight to obtain the overall average change similarity:
\begin{align}
\text{average\_change\_similarity} = \frac{\sum_{c \in \mathcal{C}} S_\text{final}(c) \, w(c)}{\sum_{c \in \mathcal{C}} w(c)}
\label{eq:change_similarity}
\end{align}

This formula prioritizes correctness on lines that were actually changed, rather than being dominated by the unchanged majority of the code, resulting in a normalized final score between 0 and 1.

\paragraph{Structural fidelity (project structure correctness).}
We extract all JSON keys/indices (i.e., folder/file structure) from target and predicted outputs into sets $K_{\text{true}}$ and $K_{\text{pred}}$ and compute precision, recall, and F1. This captures spurious files/folders (precision loss) and missing expected artifacts (recall loss). We report F1 as the structural fidelity score.
\begin{equation}\label{eq:prec_recall}
\text{Precision}=\frac{|K_{\text{true}}\cap K_{\text{pred}}|}{|K_{\text{pred}}|}
\qquad
\text{Recall}=\frac{|K_{\text{true}}\cap K_{\text{pred}}|}{|K_{\text{true}}|}
\end{equation}

\begin{equation}\label{eq:f1}
\text{F1}=\text{structural\_fidelity}=
\frac{2\,\text{Precision}\,\text{Recall}}{\text{Precision}+\text{Recall}}
\end{equation}

\textit{Human evaluation.}
We conducted a qualitative review with \textbf{four} DSL-proficient developers (\textbf{3--15} years industry experience) who use the DSL daily. They evaluated \textbf{20} outputs from the best fine-tuned model, randomly sampled and stratified across five operation types (\textsc{create}, \textsc{add} attribute, \textsc{add} finance product, \textsc{delete} attribute, \textsc{delete} finance product; 4 per type). Each reviewer independently assessed every sample, ran the existing DSL-to-code generator to detect compilation/generation errors, and could additionally submit free-form NL requests beyond the five operations for qualitative exploration. Reviewers then completed a questionnaire with nine \textbf{1--5 Likert} criteria (syntactic correctness, structural fidelity, semantic correctness, rule adherence, readability, completeness, absence of hallucinations, overall quality, usefulness) plus an open-ended question on expected time savings when creating a new market from scratch. All reviews were conducted independently to reduce group influence and capture inter-reviewer variation. Across the eight Likert-rated quality dimensions, reviewers showed high practical consistency: exact agreement (all four reviewers selecting the same score) occurred for 2/8 dimensions (25\%), and agreement within ±1 point occurred for 7/8 dimensions (87.5\%). Variation is primarily concentrated in the Completeness dimension. More detailed per-criterion distributions are shown in Figure \ref{fig:human_plot}.

\textit{Toolchain acceptance test.}
Finally, we feed selected generated DSL outputs into the existing DSL-to-code generator to verify that the outputs (i) conform to the DSL grammar and (ii) can be transformed into Java/TypeScript without generator errors. This step provides an end-to-end validation beyond surface similarity.

\textit{Replication package.} To support reproducibility, we provide a replication package containing the survey instrument and anonym-\allowbreak ized responses, analysis scripts, and the evaluation/training utilities used in this study. \footnote{https://doi.org/10.6084/m9.figshare.31423292}

\section{Results}\label{sec:results}

We report results for two models (Qwen and DeepSeek) under three configurations: \textbf{Raw} (zero-shot), \textbf{Raw+ICL} (one-shot with DSL grammar overview and a single demonstration), and \textbf{FT} (QLoRA fine-tuned). We evaluate repository-scale outputs using automatic metrics (Exact Match, Valid JSON, BLEU, Change Similarity, Structural Fidelity) and complement them with a developer study and toolchain acceptance via the DSL-to-code generator.

\subsection{RQ1: Repository-Scale Accuracy}\label{sec:rq1}
Table~\ref{tab:two_models_scores} demonstrates that repository-scale NL-to-DSL generation is feasible, but baseline accuracy varies substantially by model and setting. A first observation is that the \emph{representation itself} is not the primary bottleneck: even raw models frequently produce valid linearized project JSON (Qwen-Raw: 0.895; DeepSeek-Raw: 0.848). However, correctness at repository scale is stricter than formatting---raw models often produce outputs that are structurally plausible but not identical to the reference, reflected by low exact match (Qwen-Raw: 0.076; DeepSeek-Raw: 0.371).

\begin{table}[t]
\centering
\small
\setlength{\tabcolsep}{2.5pt}
\caption{Scores of the two evaluated models across settings.}
\label{tab:two_models_scores}
\begin{tabularx}{\columnwidth}{l l *{5}{>{\centering\arraybackslash}X}}
\toprule
Model & Setting & EM & JSON & BLEU & Change Similar. & Struct. Fidelity \\
\midrule
\multirow{3}{*}{Qwen}
  & Raw     & 0.076 & 0.895 & 0.633 & 0.372 & 0.663 \\
  & Raw+ICL & 0.171 & \textbf{1.000} & 0.898 & 0.586 & 0.958 \\
  & FT      & 0.629 & \textbf{1.000} & 0.976 & 0.834 & 0.990 \\
\midrule
\multirow{3}{*}{DeepSeek}
  & Raw     & 0.371 & 0.848 & 0.762 & 0.522 & 0.783 \\
  & Raw+ICL & 0.448 & 0.895 & 0.825 & 0.567 & 0.865 \\
  & FT      & \best{\textbf{0.657}} & \best{\textbf{1.000}} & \best{\textbf{0.986}} & \best{\textbf{0.841}} & \best{\textbf{1.000}} \\
\bottomrule
\end{tabularx}
\end{table}

Another observation is that baseline feasibility is strongly model-dependent. DeepSeek starts substantially higher than Qwen on exact match (+0.295 absolute; 0.371 vs.\ 0.076) and also leads on change similarity (0.522 vs.\ 0.372) and structural correctness (0.783 vs.\ 0.663), suggesting that some models transfer better to this DSL and project-level editing task even before adaptation.

Fine-tuning yields high repository-level accuracy for both models. DeepSeek-FT achieves 0.657 exact match, 1.000 valid JSON, 0.986 BLEU, 0.841 change similarity, and 1.000 structural correctness; Qwen-FT is close (0.629 exact match; 0.990 structural correctness). Notably, the remaining gap between near-perfect lexical overlap/structure and exact match (e.g., BLEU $\approx$ 0.98 and structure $\approx$ 1.0 vs.\ exact match $\approx$ 0.66) indicates a ``near-miss'' regime: many failures are likely due to small but critical content differences rather than missing files/folders.

Finally, across configurations, structural correctness consistently exceeds change similarity, implying that generating the correct repository \emph{shape} (files/folders) is easier than performing fully correct \emph{dependency-aware edits} across file contents, a key challenge in repository-scale DSL editing.
\begin{mdframed}[style=graybox]

\textbf{Answer to RQ1.}
LLMs can generate repository-scale DSL change-sets from a single instruction with high accuracy after adaptation: the best fine-tuned model reaches \textbf{0.657} exact match, \textbf{0.841} change similarity, \textbf{1.000} structural correctness, and \textbf{0.986} BLEU.
\end{mdframed}

\subsection{RQ2: Impact of Adaptation}\label{sec:rq2}
To quantify the benefit of adaptation, we compare \textbf{Raw} (zero-shot), \textbf{Raw+ICL} (one-shot with a compact DSL grammar overview and a single demonstration), and \textbf{FT} (QLoRA fine-tuned) for both base models as discussed in Table~\ref{tab:two_models_scores}). We focus on three outcome dimensions that matter for repository-scale DSL editing: \emph{(i) strict end-to-end correctness} (Exact Match), \emph{(ii) correctness of the requested edits} (Change Similarity), and \emph{(iii) repository shape correctness} (Structural Correctness).

\textit{Prompting (ICL) improves format and structure, especially for Qwen.}
One-shot ICL yields clear improvements over raw baselines, but its impact is uneven across models. For Qwen, ICL substantially increases structural correctness (0.663 $\rightarrow$ 0.958, +0.295) and change similarity (0.372 $\rightarrow$ 0.586, +0.214), while exact match improves only modestly (0.076 $\rightarrow$ 0.171, +0.095). For DeepSeek, ICL gains are smaller across the board (exact match: 0.371 $\rightarrow$ 0.448, +0.077; change similarity: 0.522 $\rightarrow$ 0.567, +0.045; structural correctness: 0.783 $\rightarrow$ 0.865, +0.082). This pattern suggests that a single demonstration and rule summary helps the model preserve the expected repository structure and output format, but is less reliable at enforcing the DSL-specific semantics and cross-file dependency edits that determine exact matches.

\textit{Fine-tuning delivers the largest gains in edit correctness and strict correctness.}
Fine-tuning provides the strongest overall improvements for both models, particularly on exact match and change similarity. For Qwen, FT increases exact match from 0.171 (ICL) to 0.629 (+0.458) and change similarity from 0.586 to 0.834 (+0.248), while structural correctness improves slightly further from 0.958 to 0.990 (+0.032). For DeepSeek, FT increases exact match from 0.448 to 0.657 (+0.209) and change similarity from 0.567 to 0.841 (+0.274), with structural correctness rising from 0.865 to 1.000 (+0.135). Notably, after fine-tuning both models reach near-perfect structural correctness (Qwen-FT: 0.990; DeepSeek-FT: 1.000) and high change similarity (0.834--0.841), but exact match remains lower (0.629--0.657), indicating that the remaining errors are typically small but critical deviations rather than missing files/folders.

\textit{ICL is a useful ``quick win'', but fine-tuning is required for reliable dependency-aware edits.}
ICL can move a weak baseline (Qwen-Raw) close to the desired repository structure, but the largest improvements in applying correct multi-file edits (Change Similarity) and achieving strict end-to-end correctness (Exact Match) come from supervised fine-tuning.

\begin{mdframed}[style=graybox]

\textbf{Answer to RQ2.}
Both techniques improve performance, but fine-tuning provides the dominant gains on repository-scale correctness: over Raw+ICL, it raises exact match by \textbf{+0.458} (Qwen) and \textbf{+0.209} (DeepSeek), change similarity by \textbf{+0.248} and \textbf{+0.274}, and structural correctness to \textbf{0.990--1.000}.
\end{mdframed}

\subsection{RQ3: Developer usefulness}\label{sec:rq3}

Automatic metrics do not guarantee that generated change-sets are acceptable in day-to-day maintenance workflows. We therefore evaluated developer-perceived quality and usefulness of the best-performing model (DeepSeek-FT) using (i) expert ratings and qualitative feedback and (ii) toolchain acceptance via the official DSL generator.

\begin{figure*}
    \centering
    \includegraphics[width=\textwidth]{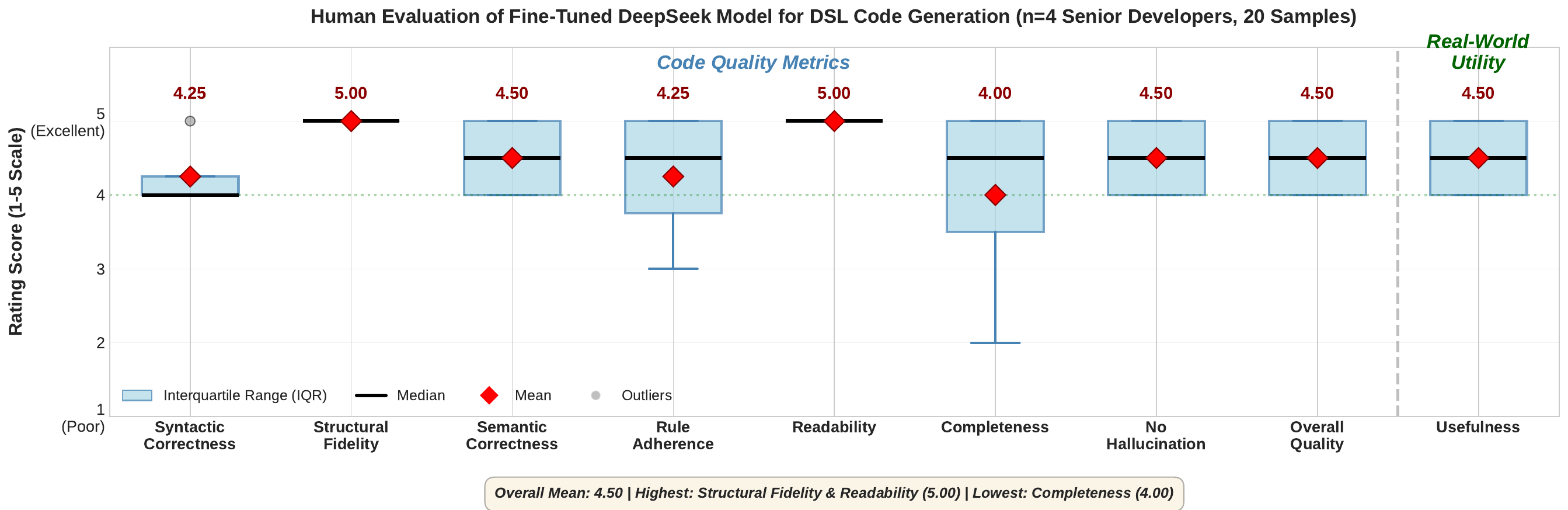}
    \caption{Distribution of human evaluation scores for the fine-tuned DeepSeek model.}

    \label{fig:human_plot}
\end{figure*}

\textit{Perceived DSL code quality.}
Four DSL-proficient senior developers reviewed 20 randomly sampled model outputs (stratified across the five operation types; 4 samples per type) and rated eight quality dimensions on a 1--5 Likert scale. Figure~\ref{fig:human_plot} summarizes the distribution of ratings. Overall, ratings indicate high practical quality: the mean score across criteria is 4.50/5, with Structural Fidelity and Readability achieving the highest mean scores (5.00/5). Semantic Correctness and Overall Quality were rated 4.50/5, while Syntactic Correctness and Rule Adherence scored 4.25/5. The lowest-rated dimension was Completeness (4.00/5), reflecting reviewer feedback that broader DSL coverage (e.g., larger markets and additional operations) would be required for full production applicability. 

\textit{Usefulness and adoption signals.}
Developers rated perceived usefulness at 4.50/5 (Figure~\ref{fig:human_plot}), and all reviewers indicated they would consider using such an LLM-based tool for daily DSL tasks.

\textit{Estimated time savings.}
Reviewers’ estimates of potential time savings for creating a new market from scratch (copying an existing market and applying required modifications) varied substantially with market complexity and similarity to existing markets. One reviewer described the expected savings qualitatively as ``a lot,'' conditional on similarity between source and target markets. Quantitatively, reviewers estimated savings of approximately \textbf{40\%}, up to \textbf{70\%} for smaller markets, and up to \textbf{80\%} under the assumption of full functional coverage across markets. In contrast, for larger markets, one reviewer estimated substantially lower savings (\textbf{5--10\%}), arguing that extensive manual correction would still be required even if the model provides a useful initial scaffold. Taken together, the feedback indicates that the tool’s perceived value is highest for market bootstrapping and repetitive boilerplate edits, while end-to-end time savings depend strongly on inference latency and training coverage of complex market configurations.

\textit{Toolchain acceptance (generator execution).}
To validate usability beyond subjective ratings, we ran the DSL-to-code generator on \textbf{20} sampled outputs (\textbf{4} per operation type). Success was perfect for \textsc{create} and both \textsc{delete} operations (\textbf{1.00}), but lower for \textsc{add} operations (\textbf{0.75} for adding an attribute; \textbf{0.75} for adding a finance product). Failures were \emph{near-miss} errors: (i) a hallucinated attribute type unknown to the generator and (ii) an unnecessary attribute added in the time-slices layer, violating DSL rules. Both required only a minimal single-line fix, but they indicate that strict toolchain compatibility is the main deployment barrier in dependency-heavy \textsc{add} scenarios.

\begin{mdframed}[style=graybox]

\textbf{Answer to RQ3.}
Developers judged the generated change-sets as highly usable in practice (overall mean \textbf{4.50/5}; usefulness \textbf{4.50/5}), with strongest performance on repository structure and readability. Remaining adoption barriers are occasional semantic/rule errors in \textsc{add} operations, plus non-functional limits such as latency and coverage of larger markets.
\end{mdframed}

\section{Discussion}\label{sec:discussion}
This section interprets the empirical findings with respect to RQ1--RQ3 and discusses implications for repository-scale NL-to-DSL generation, followed by limitations and future work.

\textit{\textbf{RQ1--RQ2: Feasibility and the effect of adaptation.}}
Across both model families, the raw (zero-shot) baselines demonstrate that the task is partially feasible but not reliable without adaptation. While both models frequently produce structurally plausible outputs in the linearized JSON representation (Valid JSON rates of 0.895 for Qwen-Raw and 0.848 for DeepSeek-Raw), strict end-to-end correctness remains low, particularly for Qwen-Raw (Exact Match 0.076) compared to DeepSeek-Raw (0.371). This gap indicates that \emph{format compliance} is not the primary difficulty; rather, the bottleneck is \emph{dependency-aware semantic editing} across files and layers (e.g., selecting valid attribute types and updating dependent locations consistently). These observations align with prior findings that general and code-focused LLMs often underperform on DSLs without task- and domain-aligned supervision \cite{gu2025, joel2024}.

Prompting (one-shot ICL) improves performance for both models, but the magnitude differs: Qwen benefits markedly (e.g., Structural Score 0.663 $\rightarrow$ 0.958; Change Similarity 0.372 $\rightarrow$ 0.586), while DeepSeek improves more modestly (Structural Score 0.783 $\rightarrow$ 0.865; Change Similarity 0.522 $\rightarrow$ 0.567). A plausible interpretation is that ICL mainly provides a \emph{structural prior}—how to format and organize repository-scale outputs—and therefore helps the weaker Qwen baseline more than the already-stronger DeepSeek baseline. However, one-shot prompting still does not yield consistently correct multi-file edits, and moving to few-shot would increase prompt length and interaction overhead, conflicting with the goal of minimal user input in practice.

Fine-tuning yields the strongest and most consistent improvements across metrics for both models. DeepSeek-FT achieves the best overall quantitative performance (Exact Match 0.657; Change Similarity 0.841; Structural Score 1.000), with Qwen-FT close behind (Exact Match 0.629; Change Similarity 0.834; Structural Score 0.990). Importantly, the remaining gap between near-perfect overlap/structure (e.g., BLEU $\approx$ 0.98 and Structural Score $\approx$ 1.0) and Exact Match ($\approx$ 0.66) suggests a \emph{near-miss regime}: many failures are likely caused by small but critical content deviations rather than missing files/folders. This is expected for repository-scale outputs where a single incorrect token (e.g., an invalid type identifier) can invalidate an otherwise correct change-set.

\textit{\textbf{RQ3: Practical utility for DSL maintainers.}}
The human evaluation provides direct evidence for practical usefulness. Four senior DSL developers rated DeepSeek-FT outputs highly across all eight quality dimensions. Perceived usefulness was also high (4.50/5), and all reviewers indicated they would consider using such a tool in daily work. Two adoption-relevant themes emerge from the open feedback. Reviewers emphasized workflow integration (IDE support) and interaction design (bulk edits), which reflects that usefulness depends not only on correctness but also on how well the tool fits existing development practices. Time-savings estimates were strongly contingent on scope and market complexity: reviewers suggested large potential savings for smaller (roughly 40--80\%), but substantially smaller savings for large markets (e.g., 5--10\%), where manual correction remains necessary.

Toolchain-based validation highlights an important practical implication: repository-scale NL-to-DSL generation can appear highly accurate while still failing under strict downstream compilation constraints. In our case, \textsc{add} operations are more fragile than \textsc{create}/\textsc{delete} because they require \emph{new} type selection and rule-consistent insertion across dependency-heavy layers, where even small semantic deviations can invalidate the change-set. For real-world adoption, this suggests pairing generation with automated safeguards, for example, generator-driven feedback for iterative repair, lightweight rule checks (e.g., type existence/allowed placements), or grammar-/constraint-aware decoding, to systematically close the remaining gap to toolchain acceptance.

\textit{\textbf{Implications for repository-scale NL-to-DSL generation.}}
Overall, linearizing repository state into a single JSON document is a viable way to teach LLMs repository-scale edits, provided the model is adapted to the DSL via fine-tuning. Structural correctness improves with both ICL and fine-tuning, and fine-tuned models reliably preserve folder/file structure. The harder challenge is semantic correctness in change-intensive cases, where outputs must satisfy both DSL grammar and domain conventions (e.g., attribute-type mappings and cross-layer dependencies). This matters for evaluation: overlap-heavy metrics over largely unchanged context (e.g., BLEU) can appear near-perfect even when strict correctness and toolchain acceptance fail, so change-focused metrics and generator-based validation are necessary complements for assessing repository-scale DSL change-sets.

\textit{\textbf{Threats to validity.}}
This study has several threats to validity. 

This is an industrial study in a specific setting, so results may not fully transfer. We report the results as an in-depth case study to provide empirical motivation and practical guidance for organizations facing similar repository-scale multi-file generation tasks.

The dataset is small and manually constructed, reflecting the practical constraint that no public dataset exists for the proprietary DSL. Limited data likely contributes to residual errors in type selection and rule adherence, especially for \textsc{add} operations. 

The human evaluation used four internal experts; while we randomized presentation to reduce ordering effects and report inter-rater reliability to assess rating consistency, the small internal panel may limit generalizability and introduce shared-context bias. The custom DSL-specific metrics capture change-level similarity and structural correctness but may penalize semantically equivalent edits placed differently; generator-based validation partly mitigates this but is also limited to the generator's notion of correctness.

\textit{\textbf{Future work.}}
Future work should evaluate generalizability and reliability at larger scales. Promising directions include (i) expanding datasets and operations to cover larger repositories and more complex edit types, (ii) incorporating constraints and validation/repair loops using grammar rules and generator feedback to improve correctness for dependency-heavy changes, and (iii) reducing context requirements via retrieval-based context selection so models can operate on larger repositories within fixed context windows. Replications on additional DSLs and domains are needed to assess external validity.

\section{Conclusion}\label{sec:conclusion}
This paper studied repository-scale NL-to-DSL generation for an industrial, Xtext-based DSL, focusing on whether a single natural-language instruction can be translated into a coherent multi-file change-set (folders, files, and contents). We operationalized repository context by linearizing each market’s DSL project into a single structured JSON document and evaluated two code-oriented LLMs (Qwen2.5-Coder and DeepSeek-Coder) under three settings: zero-shot, one-shot prompting, and parameter-efficient fine-tuning.

Across automated metrics, both prompting and fine-tuning improved performance, with fine-tuning delivering the largest gains in strict correctness and change-focused accuracy. The best fine-tuned model achieved high repository-level accuracy while preserving project structure, indicating that repository-scale DSL change-sets can be generated from a single instruction when the model is adapted to the DSL. Developer evaluation and Toolchain execution further support practical utility. Four senior DSL developers rated DeepSeek-FT outputs highly, with ceiling-level scores for structural fidelity and readability, and identified remaining adoption barriers mainly in coverage completeness, occasional rule/type selection errors, and deployment constraints such as latency. 

Overall, the paper provides evidence that linearized repository context combined with parameter-efficient fine-tuning is an effective approach for repository-scale NL-to-DSL generation. Future work should study reliability mechanisms (constraint-aware generation and validation/repair), context-selection strategies to scale to larger repositories, and replications across additional DSLs to strengthen external validity.


\bibliographystyle{ACM-Reference-Format}
\bibliography{bibfile}

\appendix

\end{document}